\documentclass[%
 reprint,
 amsmath,amssymb,
 superscriptaddress,
 aps,
 pre,
longbibliography,
]{revtex4-1}

\usepackage[utf8]{inputenc}
\usepackage[T1]{fontenc}
\usepackage{graphicx}
\usepackage{dcolumn}
\usepackage{bm}
\usepackage{amssymb}
\usepackage{amsmath}
\usepackage{tabularx}
\usepackage{hyperref} 
\usepackage{placeins}

\usepackage{blindtext}

\begin{document}
\title{Intermediate spectral statistics of rational triangular quantum billiards}

\author{\v Crt Lozej}
\affiliation{Max Planck Institute for the Physics of Complex Systems, Dresden, Germany}
\author{Eugene Bogomolny}
\affiliation{Université Paris-Saclay, CNRS, LPTMS, 91405 Orsay, France}

\begin{abstract}

 Triangular billiards whose angles are rational multiples of $\pi$ are one of the simplest examples of pseudo-integrable models with intriguing classical and quantum properties. We perform an extensive numerical study of spectral statistics of  eight quantized rational triangles, six belonging to the family of right-angled Veech triangles and two obtuse rational triangles. Large spectral samples of up to one million energy levels were calculated for each triangle which permits to  determine their spectral statistics with great accuracy. It is demonstrated that  they are of the intermediate type, sharing some features with chaotic systems, like level repulsion and some with integrable systems, like exponential tails of the level spacing distributions. Another distinctive feature of intermediate spectral statistics is a finite value of the level compressibility. 
 The short range statistics such as the level spacing distributions, and long-range statistics such as the number variance and spectral form factors were analyzed in detail. An excellent agreement between the numerical data and the model of gamma distributions is revealed.

\end{abstract}

\maketitle

\section{Introduction}
\label{sec:Intro}
The study of quantum chaos \cite{stockmannQuantumChaosIntroduction1999,haakeQuantumSignaturesChaos2001} relates concepts of classical ergodic theory to quantum systems. There are two main conjectures that connect  spectral statistical properties of quantum systems with  classical dynamical features: 
(i) The quantum chaos or Bohigas-Giannoni-Schmit conjecture \cite{casatiConnectionQuantizationNonintegrable1980,bohigasCharacterizationChaoticQuantum1984} states that generic chaotic quantum systems should have spectral statistics (or quantum statistical properties in general) that are described by an appropriate ensemble of random matrix theory (RMT). The appropriate ensemble is determined solely by unitary and anti-unitary (say time-reversal) symmetries of the system. 
(ii) The Berry-Tabor conjecture \cite{berryRegularIrregularSemiclassical1977} states that generic integrable systems are described by the Poisson statistics. Both have been tested and corroborated by an extensive number of examples, and form the foundation of quantum chaos.

However, they do not cover all possible types of dynamical systems. In this paper, we focus on the so called \emph{pseudo-integrable} models \cite{richensPseudointegrableSystemsClassical1981}. For simplicity, let us consider only two-dimensional Hamiltonian systems. Classical integrable systems of this kind are characterized by the fact that a typical trajectory will belong to a torus (i.e. a two-dimensional surface of genus 1). In contrast, the trajectories of chaotic systems will cover the entire three-dimensional surface of constant energy. In pseudo-integrable systems, the trajectories will spread over two-dimensional surfaces of genus higher than 1, which explains their name. Plane polygonal billiards whose internal angles are rational fractions of $\pi$,
\begin{equation}
    \varphi_i=\frac{n_j}{m_j}\pi, \label{eq:angles}
\end{equation}
with co-prime integers $n_j$ and $m_j$ are a characteristic example of such a system. It is proven \cite{Katok} that the genus of the surface in this case is given by 
\begin{equation}
    g=1+\frac{M}{2}\sum_j\frac{n_j-1}{m_j}, \label{eq:genus}
\end{equation}
where $M$ is the least common multiple of all the denominators $m_j$. 

In spite of their apparent simplicity, the study of polygonal billiards is notoriously challenging (see \cite{gutkinBilliardsPolygons1986, gutkinBilliardsPolygonsSurvey1996} and references therein), and analytical results are usually limited to specific cases. The mechanisms of defocusing \cite{SinaiBilliards} and focusing-defocusing \cite{bunimovichBilliardsCloseDispersing1974, bunimovichErgodicPropertiesNowhere1979} that result in chaotic dynamics in billiards are well known and require curved boundaries. Because of this, polygonal billiards are strictly non-chaotic and the well-developed methods for chaotic systems of ergodic theory do not apply. For instance, even the general periodic orbit structure and the existence of periodic orbits (see e.g., Ref.~\cite{schwartzObtuseTriangularBilliards2009} for triangles) is hard to prove. The exception are the Veech polygons \cite{veechTeichmullerCurvesModuli1989}, that possess the so-called lattice property and consequently the properties of the periodic orbits are known. 

Triangular billiards have a rich landscape of distinct dynamical regimes in their own right. Based on numerical evidence, generic triangles (all angles have irrational ratios with $\pi$) are believed to be strongly-mixing \cite{casatiMixingPropertyTriangular1999}. On the other hand, irrational triangles with one rational angle, like for instance right-triangles, have weaker ergodic properties \cite{artusoNumericalStudyErgodic1997,wangNonergodicityLocalizationInvariant2014,huangUltraslowDiffusionWeak2017} and the most recent numerical evidence suggests they are not ergodic in the Lebesgue measure \cite{zahradovaErgodicPropertiesTriangular2022,zahradovaImpactSymmetryErgodic2022,zahradovaAnomalousDynamicsSymmetric2023}. The quantum triangular billiards of these classes have been explored in \cite{lozejQuantumChaosTriangular2022}. Based on relations \eqref{eq:angles} and \eqref{eq:genus} rational triangles belong to the pseudo-integrable regime. Specifically, the list of known Veech triangles is given in Ref. \cite{hooperAnotherVeechTriangle2013}. In particular, the family of Veech right-triangles with one of the angles equal to $\pi$ divided by an integer will be of interest for this paper.

The knowledge of quantum properties of pseudo-integrable billiards is fragmentary and
includes mainly numerical calculations for billiards of simple shape: rhombus, right triangles, rectangular billiard with a barrier, etc, \cite{cheonQuantumLevelStatistics1989,shudoExtensiveNumericalStudy1993,shudoStatisticalPropertiesSpectra1994,schachnerQuantumBilliardsShape1994,bogomolnyModelsIntermediateSpectral1999,gremaudSpacingDistributionsRhombus1998,wiersigSpectralPropertiesQuantized2002}. The only quantity that is accessible to analytical derivations is the \emph{spectral compressibility} $\chi$, which determines the growth of the variance \cite{berrySemiclassicalTheorySpectral1985} of the of number of levels in an interval of length $L$
\begin{equation}
    \left\langle (N(L)-L)^2 \right\rangle  \underset{L \rightarrow \infty}{\sim} \chi L,
\end{equation}
where $N(L)$ is normalized such that its mean value equals $L$ and the angled brackets denote an averaging over a small energy window. The compressibility distinguishes between chaotic $\chi=0$ and integrable models $\chi=1$. The calculation of the compressibility is done by the summation over classical periodic orbits in the diagonal approximation \cite{berrySemiclassicalTheorySpectral1985}. In the particular case of the above-mentioned Veech right-triangles with angles $\pi/m$, the compressibility is given by \cite{bogomolnyPeriodicOrbitsContribution2001}
\begin{equation}
    \chi=\frac{m+\epsilon(m)}{3(m-2)}, \label{eq:Veechcompress}
\end{equation}
where depends on the factors of $m$ such that $\epsilon(m)=0$ for odd $m$, $\epsilon(m)=2$ for even $m$ but $m\not\equiv0\mod{3}$, and $\epsilon(m)=6$ for $m\equiv0\mod{6}$.    Similarly, analytical results in barrier billiards show, $\chi=1/2$ regardless of the barrier height \cite{bogomolnyBarrierBilliardRandom2021,bogomolnyRandomMatricesAssociated2022}. As we see, the spectral properties of pseudo-integrable systems can subtly depend on structural details. The fact that for these models $0 < \chi < 1$ is indicative that spectral statistics of such billiards are different from those of both chaotic and integrable models, but share similarities with both. This is why they are referred to as \emph{intermediate spectral statistics}. The most well known example is the Semi-Poisson statistics \cite{bogomolnyModelsIntermediateSpectral1999} that may be obtained by taking only every second level of a Poissonian spectrum. Incidentally, this corresponds to $\chi=1/2$.  

Numerical observations of intermediate spectral statistics have established the following properties: (i) Level repulsion at small distances, as in the standard random matrix ensembles. (ii) Exponential tails of nearest-neighbor spacing distributions, as in the Poissonian case. (iii) Non-trivial value of the spectral compressibility. (iv) Multi-fractal dimensions of eigenfunctions \cite{bogomolnyStructureWaveFunctions2004, bogomolnyFormationSuperscarWaves2021}.
This type of statistics have first been observed at metal–insulator transition point in the Anderson model \cite{altshulerRepulsionEnergyLevels1988, shklovskiiStatisticsSpectraDisordered1993} and in a variety of other dynamical systems including the already mentioned billiards, but also in neutrino billiards \cite{dietzSemiPoissonStatisticsRelativistic2023}, quantum maps \cite{giraudIntermediateStatisticsQuantum2004}, short-range plasma models \cite{bogomolnyModelsIntermediateSpectral1999,bogomolnyShortrangePlasmaModel2001} and models of structured random matrices \cite{bogomolnySpectralStatisticsRandom2020, bogomolnyStatisticalPropertiesStructured2021}.

The principal results of  the paper are:  (i) the spectral statistics of rational triangular quantum billiards are of the intermediate type and (ii) the correlation functions  are well described   by  simple  gamma distribution formulas.  
These conclusions are based   on  numerical computations of  large spectra of up to 1 million eigenenergies of eight triangles, namely six Veech right-triangles and two non-Veech obtuse triangles, with subsequent analysis of  nearest-neighbour and higher orders level spacing distributions, the number variance,  the spectral form factor, and level compressibility   for all these  models. 

The plan of the paper is the follows. In Sec. \ref{sec:Definitions} the quantum billiard problem, the geometries of the triangular billiards and the spectral statistics considered in this study are defined. In Sec. \ref{sec:Spectral statistics} the numerical results are presented and analyzed. In Sec. \ref{sec:Conclusion} the conclusions are presented and discussed. Appendix \ref{AppendixA} offers a heuristic explanation for the success of the gamma distributions by using a toy model. In Sec. \ref{AppendixB} some more complicated two-parameter fitting distributions are presented.

\section{Definitions}
\label{sec:Definitions}
This section gives all the relevant definitions of the dynamical system, quantities, and their relationships considered in this paper, and introduces notation.

\subsection{Quantum billiards}
Dynamical billiards are archetypical models of both classical and quantum chaos. In a two-dimensional quantum billiard problem, one considers a quantum particle trapped inside a region $\mathcal{B} \subset \mathbb{R}^2$ known as the billiard table.  The eigenfuncitons $\psi_n(x)$ are given by the solutions of the Helmholtz equation 
\begin{equation}
    \left(\nabla^{2}+k_n^{2}\right)\psi_n(x)=0 
\end{equation}
with certain boundary conditions (b.c.).  In this study,  only triangular regions and Dirichlet b.c. are considered.  In means that  $\psi_n|_{\partial \mathcal{B}}=0$, with eigenenergies $E_n=k_n^2$, where $k_n$ is the wavenumber of the $n$-th eigenstate. 

The very efficient scaling method, devised by Vergini and Saraceno \cite{verginiCalculationScalingHighly1995,verginiEstudioCuanticoSemiclasico1995} and extensively studied by Barnett \cite{barnettDissipationDeformingChaotic2001}, with a corner adapted Fourier-Bessel basis \cite{barnettQuantumMushroomBilliards2007}, allows us to compute very large spectra of the order of $10^6$ states. The implementation is available as part of \cite{lozejQuantumBilliards} and is the same as used in \cite{lozejQuantumChaosTriangular2022}. 

The spectral staircase function counts the number of eigenstates (or modes) up to some energy $N(E) := \#\{n|E_n<E\}$.
The asymptotic mean of the spectral staircase for billiards is given by the well known generalized Weyl's law \cite{baltesSpectraFiniteSystems1976}

\begin{equation}
    N_{\mathrm{Weyl}}(E) = (\mathcal{A}E-\mathcal{L}\sqrt{E})/4\pi + \sum_i{\frac{\pi^2-{\varphi_i}^2}{24 \pi {\varphi_i} }},
\end{equation}
where $\mathcal{A}$ is the area of the billiard and $\mathcal{L}$ the circumference and $\varphi_i$ are the internal angles. To compare the universal statistical fluctuations for different billiards, it is convenient to unfold the spectra. This is done by inserting the numerically computed billiard spectrum into Weyl's formula $e_n := N_{\mathrm{Weyl}}(E_n)$. The resulting unfolded spectrum ${e_n}$ has a uniform mean level density equal to one.

\subsection{Geometry}
In this study, the spectra of eight triangular quantum billiards with all angles having rational ratios with $\pi$ are considered. Of these, six were taken from the family of Veech right triangles, with angles $(\frac{\pi}{m},\frac{(m-2)\pi}{2m},\frac{\pi}{2})$ for $m\geq4$. The triangles from this group are labeled as $V_m$. Only six triangles $V_5$, $V_7$, $V_8$, $V_9$, $V_{10}$, $V_{18}$ are investigated. The triangles $V_4$ and $V_6$ are integrable and thus not interesting for this study. 

As explained in the introduction, the periodic orbit structure of the Veech triangles is known and therefore semiclassical techniques may be used to gain analytical insight into their spectral statistics, namely Eq.~\eqref{eq:Veechcompress} for the level compressibility. 

\begin{figure}
  \centering
  \includegraphics[]{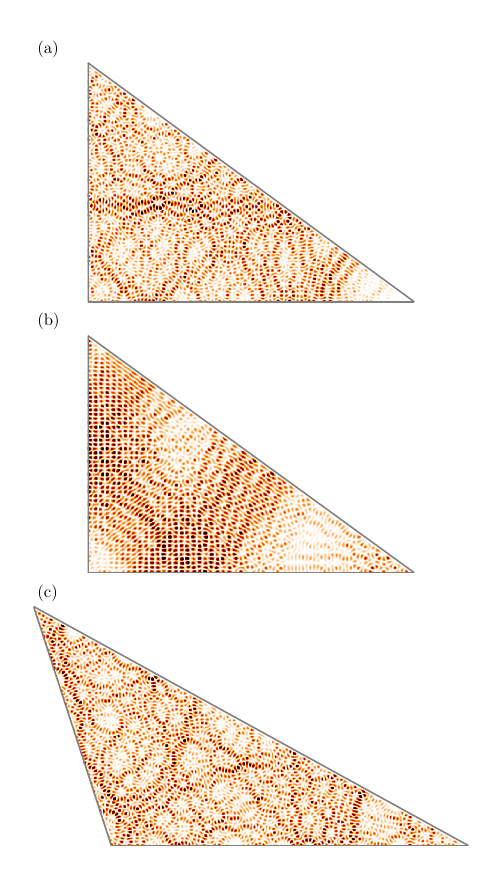}
  \caption{Examples of eigenstates of rational triangular billiards. We show the probability distribution $|\psi_n|^2$ for (a) the Veech triangle $V_5$, generic eigenstate $n=3342$ at $k=250.0099$, (b) superscar eigenstate $n=2430$ at $k=213.6317$ and (c) obtuse triangle $T_2$, generic eigenstate $n=3624$ at $k=250.0095$.} 
  \label{states}
\end{figure}

To expand the scope of our study to more general rational triangles two non-Veech obtuse triangles with angles  
$(\frac{2\pi}{15},\frac{4\pi}{15},\frac{3\pi}{5})$ and 
$(\frac{2\pi}{25} ,\frac{6\pi}{25},\frac{3\pi}{5})$ are also considered. These triangles are labeled, respectively,  by $T_1$ and $T_2$. In these cases, the classical periodic orbit structure is not known. The height, measured from the bottom side of the triangles, is fixed to $h=1$ in all cases. As an illustration, Fig. \ref{states} shows typical eigenstates of the triangles $V_5$ and $T_2$ as well as a superscar state of $V_5$. For more information about superscars the reader is referred to Refs. \cite{bogomolnyStructureWaveFunctions2004, bogomolnyFormationSuperscarWaves2021}.

\subsection{Level spacings}
The distributions of level spacings are the most commonly considered spectral statistics. Let $P_n(s)$ be the $n$-th nearest neighbor spacing distribution, that is, the probability density of the energy distances $s$ between two levels that have $n$ levels between them. For $n=0$ this is the nearest-neighbor level spacing distribution, which is the most studied. The distributions follow the normalization conditions
\begin{equation}
\int^\infty_0{P_n(s)ds=1},\quad  \quad  \int^\infty_0{sP_n(s)ds=1 + n}. 
\label{eq:normalization}
\end{equation} 
There are several well-supported conjectures that relate level spacing distributions of quantum mechanical dynamical systems to random matrix models. The Bohigas-Giannoni-Schmit conjecture \cite{casatiConnectionQuantizationNonintegrable1980,bohigasCharacterizationChaoticQuantum1984} states the spectral statistics of chaotic models will follow the statistics of Gaussian random matrices and the Berry-Tabor conjecture \cite{berryLevelClusteringRegular1977} states that integrable models follow Poissonian statistics. In chaotic systems with time reversal symmetry,   the Wigner surmise gives an excellent approximation for the nearest-neighbor spacing distribution
\begin{equation}
P_W(s) = \frac{\pi}{2} s \exp \left( - \frac{\pi}{4} s^{2} \right). \label{eq:WignerP}
\end{equation} 
Higher-order spacing surmises are given in \cite{raoHigherorderLevelSpacings2020}. In the integrable (Poissonian) case the distribution is exponential
\begin{equation}
P_I(s) =  \exp \left( -s\right). \label{eq:PoissonP}
\end{equation} 
The behavior of $P(s)$ as $s\rightarrow0$ is a notable and distinguishing feature. We see $P(s)\propto s^\beta$ ,  for small $s$, with $\beta=1$ in the chaotic and $\beta=0$  in the integrable case.  The energy levels in the chaotic case tend to form a gap (are repulsed) between each other, induced by correlations in the energy spectrum. In the integrable case, the levels are uncorrelated and have no level repulsion. The exponent $\beta$ is called the level repulsion exponent. However, interesting dynamical regimes between chaos and integrability also exist, for instance dynamical localization (See for instance \cite{izrailevSimpleModelsQuantum1990, batisticDynamicalLocalizationChaotic2013, batisticLevelRepulsionExponent2018,  lozejSpectralFormFactor2023}).  These can be modelled by using the Brody distribution \cite{brodyStatisticalMeasureRepulsion1973}, which interpolates the two regimes 
\begin{equation}
P_B(s) = a s^{\beta} \exp \left( - b s^{\beta +1} \right)\label{eq:BrodyP}
\end{equation} 
where the normalization constants \eqref{eq:normalization}  with $n=0$ are given by $a = (\beta +1 ) b,$ and $b  = \left( \Gamma \left( \frac{\beta +2}{\beta +1}\right) \right)^{\beta +1}$, where  $\Gamma(x)$ is the gamma function.  We see that both the level repulsion and the tail of the distribution change as we interpolate from $\beta=0$  to $\beta=1$ . However, as we will later confirm with the numerics, the level spacings of the rational triangles have slightly different characteristics, namely (i) Level repulsion at small $s$ and (ii) a purely exponential  tail at large $s$. Because they share characteristics of both chaotic and integrable spectra, they are known as \textit{intermediate spectral statistics}. The most well known example is the Semi-Poisson distribution,
\begin{equation}
P_{SP}(s) = 4s \exp \left( -2s\right). \label{eq:SemiPoissonP}
\end{equation} 
The main model for the level spacings of the rational triangles will be a normalized version of the family of  gamma distributions.  The probability density function for the $n$-th level spacing is given by
\begin{equation}
P_n(s) = a_n s^{\gamma_n} \exp \left( -b_n s\right), \label{eq:gammaP}
\end{equation} 
where $a_n$ and  $b_n$  are obtained from the normalization conditions \eqref{eq:normalization},
\begin{equation}
a_n=\frac{1}{\Gamma(\gamma_n+1)}\left(\frac{\gamma_n+1}{n+1}\right)^{\gamma_n+1},\quad  \quad  b_n=\frac{\gamma_n+1}{n+1}. 
\end{equation} 
The parameter $\gamma_n$ will depend on the order of the level spacing $n$. It appears that  a linear dependence
\begin{equation}
\gamma_n = p n +\gamma_0
\label{eq:gammadep}
\end{equation} 
where $\gamma_0$ and $p$  are fitting parameters is consistent with the numerical data. Setting $\gamma_0=0$ and $p=1$ corresponds to the Poisson model and $\gamma_0=1$ and $p=2$  to Semi-Poisson.  The gamma model is not new and has been used previously to describe spectra of structured random matrices \cite{bogomolnyStatisticalPropertiesStructured2021}, where the parameters are related to the “zero modes” of the matrices that is, the number of parameters whose variation does not remove the eigenvalue degeneracy. In appendix \ref{AppendixA} a toy model that offers a heuristic explanation for the success of the gamma model is presented.  

\subsection{Two-point correlations}
The two-point correlation function $R_2(s)$ is the probability that two levels are separated by the distance $s$.  Since there can be any number of levels in-between, it is equal to the sum over all orders of the spacing distributions
\begin{equation}
    R_2(s) = \sum^\infty_{n=0}{P_n(s)}. 
\label{eq:twoponit}
\end{equation}
There are many interesting spectral statistics related to the two-point correlation function. In this paper, we will focus on the number variance and the spectral form factor. The number variance is the local variance of the number of levels in an interval of length $L$,  given by
\begin{equation}
    \Sigma^2(L,e):=\left\langle (N(L,x)-L)^2 \right\rangle_{e,w} \quad L>0
\end{equation}
where $N(L,x) = N(x+L/2)-N(x-L/2)$ is the number of unfolded energy levels $e_n$ in the interval $[x-L/2,\,x+L/2]$. The brackets $\langle ... \rangle_{e,w}$ denote a local average around the central energy $e$ and window width $w$, so that $x\in[e-w/2,\,e+w/2]$. The number variance is related to the two-point correlations via the integral,
\begin{equation}
    \Sigma^2(L) = L - 2\int^L_0(L-s)(1-R_2(s))ds.
\end{equation}
Particular interesting is  the long range limit of the number variance
\begin{equation}
    \Sigma^2(L)  \underset{L \rightarrow \infty}{\sim} \chi L  
\end{equation}
where the proportionality coefficient $\chi$  is called the \textit{level compressibility}. Therefore,  the compressibility can be defined as
\begin{equation}
    \chi = \underset{L \to \infty}{\lim} \frac{\Sigma^2(L)}{L}. 
\end{equation}
This yields $\chi = 0$ for the Gaussian ensembles of random matrices  and $\chi = 1$ in the Poisson case. For the intermediate spectral statistics, it is argued \cite{altshulerRepulsionEnergyLevels1988, shklovskiiStatisticsSpectraDisordered1993}  that  $0<\chi<1$.  Since the number variance can be expressed from the two-point correlation function, one may use the  relation  \eqref{eq:twoponit},  the model assumptions  \eqref{eq:gammaP}, and \eqref{eq:gammadep}  to compute the spectral compressibility for the gamma model  (see Ref. \cite{bogomolnyStatisticalPropertiesStructured2021} for a complete derivation). The end result is simply 
\begin{equation}
    \chi = \frac{1}{p}.
\end{equation}
The spectral form factor (SFF) is the Fourier transform of the two-point correlation function
\begin{equation}
    K(t) = \int^\infty_{-\infty} R_2(s) e^{2\pi i t s} ds. \label{eq:sffdef}
\end{equation} 
The compressibility may also be expressed as the limit 
\begin{equation}
    \chi = \underset{t \to 0}{\lim} K(t). \label{eq:sffcompress}
\end{equation}
The form factor can be expressed formally through  the spectrum as follows
\begin{equation}
    K(t) = \left\langle \left|\sum_n^N  \mathrm{exp}(2 \pi i e_n t) \right|^2\right\rangle, \label{eq:sff}
\end{equation} 
where the sum goes over the unfolded energy levels, and $\langle\cdots\rangle$ represents an average over an ensemble of similar systems or a moving time average as discussed below at the end of the section. The form factor is a very sensitive measure of quantum chaos due to its very distinct behavior in different dynamical regimes. In the Poissonian (integrable) case $K(t) = 1$. In the chaotic (GOE) case it is given by the formula \cite{mehtaRandomMatrices2004}
\begin{equation}
K_\mathrm{GOE}(t) =\begin{cases} 2t-t\mathrm{ln}(2t+1) &t < 1 \\
                                2-t\mathrm{ln}(\frac{2t+1}{2t-1}) &t > 1 \end{cases}.
\end{equation} 
From this expression it follows that   $K_\mathrm{GOE}(t\rightarrow0) = 0$. 

In the case of intermediate spectral statistics, we may expect $K(t)$ will approach a nontrivial value $0<\chi<1$ as $t\rightarrow 0$. Following Ref.~\cite{bogomolnyStatisticalPropertiesStructured2021}, the Fourier transform \eqref{eq:sffdef} may be evaluated by introducing the Laplace transform  
\begin{equation}
K(t) = 1 + \mathrm{Re}\,g(2\pi i t). \label{eq:sffLaplace}
\end{equation} 
where $g$ is defined as
\begin{equation}
g(\tau) = \sum^\infty_{n=0}{g_n(\tau)}, \label{eq:LaplaceSum}
\end{equation} 
that is the Laplace transform of the sum \eqref{eq:twoponit} and
\begin{equation}
g_n(\tau) = \int^\infty_{0} P_n(s) e^{- \tau s} ds.
\end{equation}
Using the gamma distributions \eqref{eq:gammaP} and the linear dependence \eqref{eq:gammadep} a technical derivation (see Ref.  \cite{bogomolnyStatisticalPropertiesStructured2021}) yields the expression
\begin{equation}
g(\tau) = \frac{\left(1 + \frac{\tau}{p} \right)^{p-\gamma_0-1}\hfill}{\left(1 + \frac{\tau}{p} \right)^{p}-1 \hfill}\exp{\left(-\frac{\tau(p-\gamma_0-1)}{p+\tau}\right)}, \label{eq:LaplaceFun}
\end{equation} 
which can be used to evaluate \eqref{eq:sffLaplace}. Taking the limit $t\rightarrow0$ produces $\chi=1/p$, which is, of course, consistent with the derivation from the number variance. 

    Finally, we shall make a few comments regarding the numerical evaluation of the spectral form factor. It is known the SFF is not a self averaging quantity \cite{prangeSpectralFormFactor1997} and exhibits erratic fluctuations with time. This means a separate averaging must be performed, represented by $\langle\cdots\rangle$. This is commonly an average over different realizations when considering random matrices or disordered systems. For clean single-body systems such as billiards, instead it is necessary to perform a moving time average to smooth out the fluctuations \cite{delonNOJetCooled1991,altCorrelationholeMethodSpectra1997}. The procedure is exactly as used in previous papers \cite{lozejQuantumChaosTriangular2022,lozejSpectralFormFactor2023}, where we refer the reader for further details. Furthermore, when considering the $t\rightarrow0$ limit in definitions such as \eqref{eq:sffcompress} it is implicitly assumed the limit $N \rightarrow \infty$ is taken first, i.e., the whole infinite  spectrum is considered, which is not possible in the numerics. Therefore, the limits are necessarily inverted and formally the form factor diverges at $t=0$. This can be seen e.g. by taking Eq. \eqref{eq:sff} and considering the limit $\underset{N \rightarrow \infty}{\lim} K(0) = N \rightarrow \infty$.  This drawback can be avoided by decomposing the form factor into the connected and disconnected parts (stemming from the connected and disconnected two-point correlation functions).  The disconnected part is given by the diagonal terms from Eq. \eqref{eq:sff} and depends solely on the density of states (see 
    Ref.~\cite{winerHydrodynamicTheoryConnected2022} for more details). Considering only the connected part of the spectral form factor by subtracting the disconnected part eliminates the divergences in the numerical calculations. Hence, only the connected parts of the spectral form factors are shown in the numerical results. 

\section{Numerical Results}\label{sec:Spectral statistics}
The spectral samples were produced by the scaling method as implemented in \cite{lozejQuantumBilliards}.  The scaling method computes the states in some small, finite spectral interval. The final spectral sample is a composite of many small overlapping spectral samples, where we try to identify which of the levels in the overlap interval belong to the same eigenstates. Because of the finite precision and numerical errors in the computation of the individual levels this is not always possible, and some levels are missed, while some may be counted twice. To omit the non-universal aspects of the low-lying eigenstates, we start collecting the levels from the 10000-th onward. The samples for the Veech triangles contain $10^6$ consecutive levels, the exception being  $V_{18}$ where we start with the $10^5$-th level and gather only $8\times10^5$ levels. The remaining two samples contain  $6\times10^5$ levels for $T_1$ and  $10^6$ levels for $T_2$. The number of mistakes is  less than 100 in all cases which should have no significant impact on the results. 

\subsection{Level spacings}\label{sub:Level spacings} 
\begin{figure}
  \centering
  \includegraphics[]{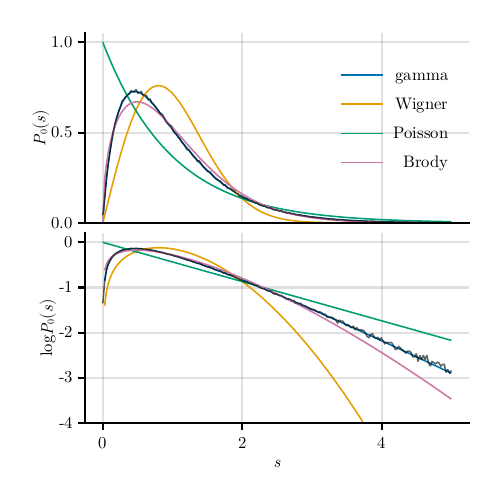}
  \caption{Nearest neighbor level spacing distributions in triangle $V_5$. The numerical results are shown in gray. The colored curves show the model predictions.} 
  \label{level_spacings}
\end{figure}
We will start with examining the level spacing distributions. In Fig. \ref{level_spacings}  we show the nearest neighbor level spacings for the Veech triangle $V_5$ compared to the analytical models in the linear (top) and log-linear scales. As expected, we observe intermediate spectral statistics, as neither the Poissonian \eqref{eq:PoissonP}  nor Wigner-Dyson \eqref{eq:WignerP} models fit the data. The Brody distribution \eqref{eq:BrodyP} is better and captures the level repulsion, but clearly misses the top and the tail of the distribution.  The gamma distribution fit is clearly the best and fits the data nearly perfectly. It is evident from the log-linear plot that the tail of the distribution is indeed exponential.  
\begin{figure}
  \centering
  \includegraphics[]{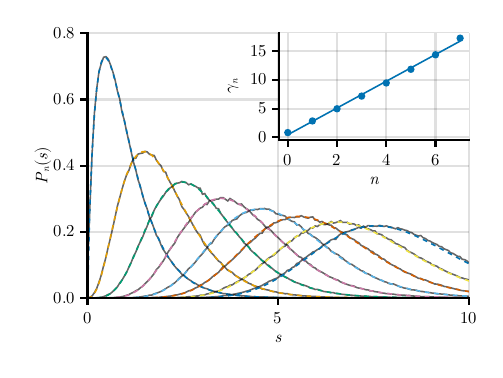}
  \caption{Level spacing distributions up to order $n=7$ for the triangle $V_5$. The numerical data is shown in gray. The dashed colored curves show the fitted gamma distributions \eqref{eq:gammaP}. The Insert shows the linear dependence of the fitted parameters $\gamma_n$ together with the best fitting line \eqref{eq:gammadep}.} 
  \label{level_spacings_higher_1}
\end{figure}
In Fig. \ref{level_spacings_higher_1} we show the level spacings for order up to $n=7$ in the same triangle $V_5$.  We compare the data with the best fitting gamma distribution for each order. Again, the distributions fit the data extremely well. The Insert shows the values of the fitted $\gamma_n$ as a function of $n$. A linear dependence of type \eqref{eq:gammadep}  with parameters $\gamma_0 = 0.75$  and  $p=2.91$  approximates the data well.  
\begin{figure}
  \centering
  \includegraphics[]{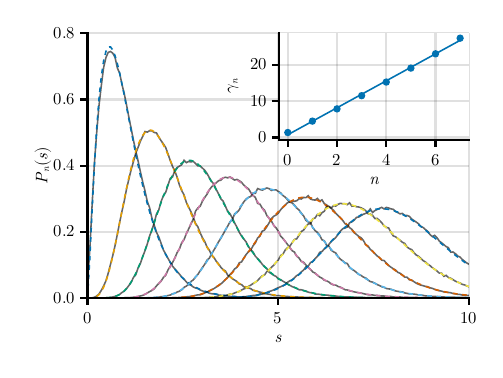}
  \caption{Same as Fig. \ref{level_spacings_higher_1}, but for the obtuse triangle $T_1$.} 
  \label{level_spacings_higher_2}
\end{figure}
In Fig. \ref{level_spacings_higher_2}  the same quantities for the rational triangle  $T_1$ are shown. The fits are of similar quality with  $\gamma_0 = 1.14$  and  $p=4.62$.  We repeat the procedure (the best fitting $\gamma_n$ for $n \leq 4$ is given in table \ref{tab:fitparams})  for the other triangles, and then obtain values for $\gamma_0$  and  $p$ from the linear fits.  In Fig. \ref{level_spacings_higher}  we show $P_n(s)$ up to $n=4$ in the log-linear scale, for all the triangles considered in this study. Instead of fitting the gamma distribution for each order separately, we use the parameters  $\gamma_0$  and  $p$ extracted from the previous step and use the linear relationship to determine the next $\gamma_n$ in the sequence. Comparing the data to the model curves, we see very good agreement between the model and the numerics. The largest deviations occur in the tails of nearest neighbor spacings, $P_0(s)$ since the fitted linear sequence sometimes slightly underestimates the value of $\gamma_0$ compared to the best fitting parameter. These small deviations would hardly be visible in the linear plot, and we may conclude the model is a very good approximation despite its simplicity. Some more complicated two-parameter extensions for fitting distributions are explored in Appendix \ref{AppendixB}. It is interesting that the gamma model works equally well for the Veech triangles and the other two rational triangles $T_1$ and $T_2$ so it  might be expected that it holds for a typical rational triangle billiard, and, conjecturally,  for pseudo-integrable systems in general. The slopes of the fitted lines give an estimate of the compressibility as $\chi_\gamma = 1/p$. The values are given in table \ref{tab:compresibility} and compared with the other methods of estimation. 

\begin{table}[]

\centering\caption{Model parameters and spectral compresibilities for all triangles. The linear gamma model parameters Eq. \eqref{eq:gammadep} are the intercept $\gamma_0$ and slope $p$. The compresibilites are obtained from the spectral form factor $\chi_K$ (coinciding with $\chi_{\gamma}=1/p$) Fig. \ref{sff}, the slope of the number variance $\chi_{\Sigma^2}$ Fig. \ref{number_variance_all} and the semiclassical periodic orbit computation (only for the Veech triangles) \eqref{eq:Veechcompress}.}{\label{tab:compresibility}}
\begin{tabular}{>{\centering}p{1.2cm}>{\centering}p{1.2cm}>{\centering}p{1.2cm}>{\centering}p{1.2cm}>{\centering}p{1.2cm}>{\centering}p{1.2cm}}
\multicolumn{6}{p{8.6cm}}{}\tabularnewline
\hline 
\hline 
label & $\gamma_0$ & $p$ & $\chi_K$ & $\chi_{\Sigma^2}$ & $\chi_{PO}$ \tabularnewline
\hline 
\hline 
$V_5$ & 0.75 & 2.15 &  0.46 &  0.35 &  0.55\tabularnewline
$V_7$ & 0.95 & 2.81 & 0.36 &  0.28 &  0.47\tabularnewline
$V_8$ & 0.76 &  2.07 & 0.48 & 0.36 &  0.55 \tabularnewline
$V_9$ & 1.08 & 3.25 & 0.31 & 0.25 &  0.43 \tabularnewline
$V_{10}$ & 0.78 & 2.40 & 0.42 & 0.29 &  0.50\tabularnewline
$V_{18}$ & 1.10 &  3.07 & 0.33 & 0.26 &  0.50\tabularnewline
$T_1$ & 1.14 &  3.48 & 0.29 & 0.20 &  N/A\tabularnewline
$T_2$ & 1.42 &  4.39 & 0.23 & 0.19 &  N/A\tabularnewline

\hline 
\end{tabular}
\end{table}

\begin{table}[]

\centering\caption{The best fitting parameters $\gamma_n$ for the gamma distributions describing the level spacing distributions $P_n(s)$ up to $n=4$.}{\label{tab:fitparams}}
\begin{tabular}{>{\centering}p{1.2cm}>{\centering}p{1.2cm}>{\centering}p{1.2cm}>{\centering}p{1.2cm}>{\centering}p{1.2cm}>{\centering}p{1.2cm}}
\multicolumn{6}{p{8.6cm}}{}\tabularnewline
\hline 
\hline 
label & $\gamma_0$ & $\gamma_1$ & $\gamma_2$ & $\gamma_3$ & $\gamma_4$ \tabularnewline
\hline 
\hline 
$V_5$ & 0.84 & 2.86 &  4.97 &  7.17 &  9.46\tabularnewline
$V_7$ & 1.12 & 3.66 & 6.41 & 9.31  &  12.35\tabularnewline
$V_8$ & 0.82 &  3.66 & 6.41 & 6.96 &  9.10 \tabularnewline
$V_9$ & 1.29 & 4.20 & 7.38 & 10.74 &  14.26 \tabularnewline
$V_{10}$ & 0.92 & 3.10 & 5.47 & 7.93 &  10.52\tabularnewline
$V_{18}$ & 1.27 &  4.07 & 7.07 & 10.25 &  13.53\tabularnewline
$T_1$ & 1.36 &  4.50 & 7.88 & 11.51 &  15.27\tabularnewline
$T_2$ & 1.69 &  5.64 & 9.95 & 14.52 &  19.18\tabularnewline

\hline 
\end{tabular}
\end{table}

\begin{figure*}
  \centering
  \includegraphics[]{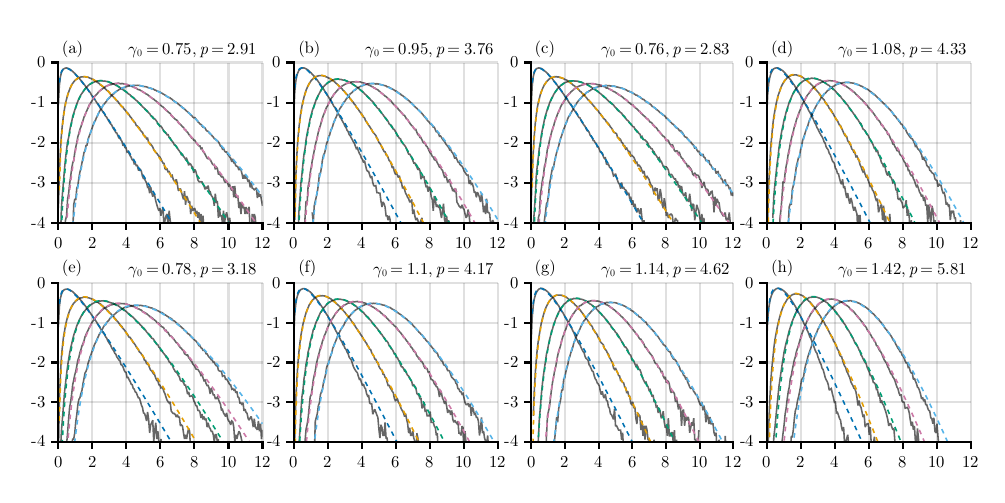}
  \caption{Level spacing distributions in the decadic logarithmic scale $\mathrm{logP_n(s)}$ up to order $n=4$, for all triangles considered in the study. The numerical data is shown in gray and the dashed colored curves show the gamma distributions \eqref{eq:BrodyP} with parameters $\gamma_n$ given by the linear relation \eqref{eq:gammadep}. The linear parameters are obtained by fitting the line as shown in Fig. \ref{level_spacings_higher_1} and noted at the top right of each panel. The panels show the triangles (a) $V_5$, (b) $V_7$, (c) $V_8$, (d) $V_9$, (e) $V_{10}$, (f) $V_{18}$, (g) $T_1$, (h) $T_2$.} 
  \label{level_spacings_higher}
\end{figure*}

\begin{figure*}
  \centering
  \includegraphics[]{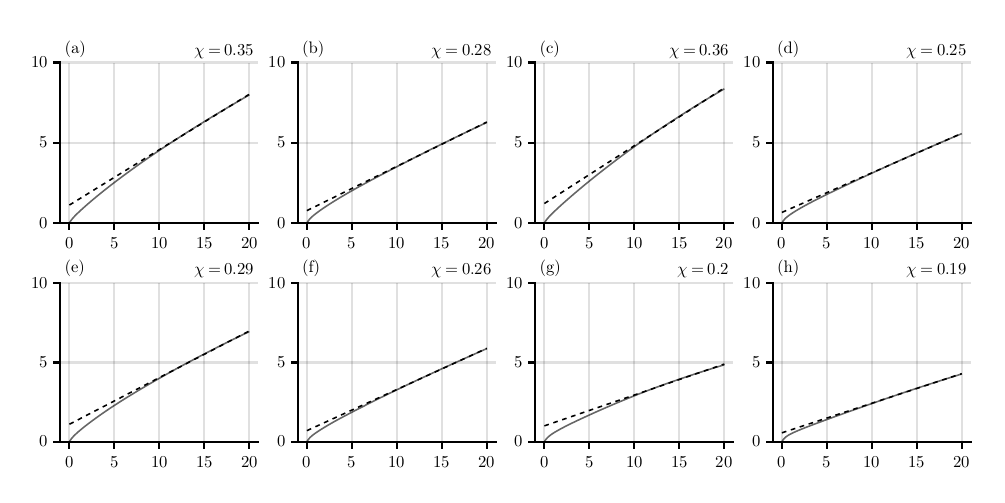}
  \caption{The number variance $\Sigma^2(L)$ for all triangles considered in the study. The data is shown in gray. The black dashed line shows the slope that best fits the data, giving the compressibility $\chi$. The sequence of triangles is the same as in Fig. \ref{level_spacings_higher}.} 
  \label{number_variance_all}
\end{figure*}

\begin{figure*}
  \centering
  \includegraphics[]{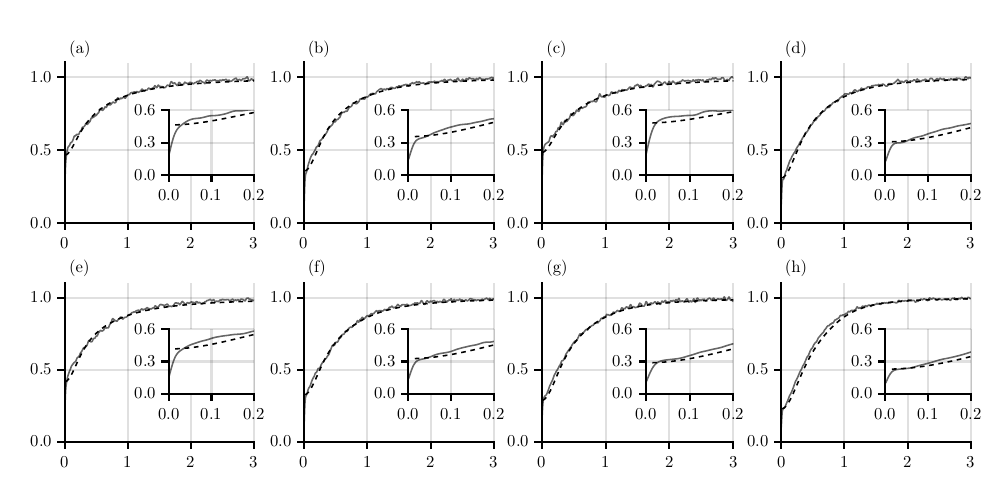}
  \caption{Connected spectral form factors $K(t)$, for all triangles considered in the study. The data is shown in gray. The black dashed curve shows the model obtained by inserting the expression \eqref{eq:LaplaceFun} into Eq. \eqref{eq:sffLaplace}. The parameters $p$ and $\gamma_0$ are the same as in Fig. \ref{level_spacings_higher}. The sequence of triangles is the same as in Fig. \ref{level_spacings_higher}.} 
  \label{sff}
\end{figure*}

\subsection{Two-point correlations}\label{sub:Number variance} 

This section is devoted to the discussion of  the results for the number variance and spectral form factors. Let us first examine the number variance. Fig. \ref{number_variance_all} shows the number variance for each triangular billiard. The qualitative behavior is similar in all cases and the number variance reaches   a seemingly linear regime at $L\approx10$. This corroborates with the  expectations of intermediate type statistics in rational billiards. The  fit of straight lines permits to determine  the slopes which determine the  compressibility $\chi_{\Sigma^2}$. The results are given in table \ref{tab:compresibility}. Since the spectra are finite, the number variance will eventually saturate and oscillate around the saturation plateau (see \cite{backerSpectralStatisticsQuantized1995} and references therein). Because of this, it is not easy to extrapolate if the asymptotic regime has been reached. The extracted compressibilities are therefore less reliable and may be seen as only a rough estimate. 

The connected  spectral form factors (see discussion at the end of Sec. \ref{sec:Definitions}) are plotted in Fig. \ref{sff}. It is again evident that the spectral statistics  are of the intermediate type as $K(t)$ approaches a finite value as $t\rightarrow0$,  in all triangles under consideration. The numerical results are compared to the analytical curves obtained by inserting the Laplace transform of the correlation function based on the gamma model \eqref{eq:LaplaceFun} into Eq. \eqref{eq:sffLaplace}. The function has two parameters $p$ and $\gamma_0$ that are given by the slope and the intercept of the linear relation \eqref{eq:gammadep}. In order to check the consistency of the model,  the parameters to the numerical form factors are not fitted but instead   the parameters obtained in fitting the level spacing distributions, given in table \ref{tab:compresibility}, are used. We see the analytical curves fit the data quite well, confirming that the parameters give consistent results. The insets show the small $t$  behavior, where we observe an oscillation resulting in a sudden drop, as $t\rightarrow0$ in the numerical data. Very similar behavior was observed for generic right-triangles in Ref. \cite{lozejQuantumChaosTriangular2022}. This is likely a consequence of the finite-size of the spectral sample, as the sudden drop tends to move further towards 0 as the sample size is increased and can be considered a finite size effect. Since, the analytical curves are based on the same fitting parameters as the level spacing distributions, they imply the same values for the compressibility, namely $\chi=1/p$. 

The complete results for the compressibilities are gathered in table \ref{tab:compresibility}. We observe the values, extracted from the level spacings (equivalently spectral form factors) and the ones extracted from the number variances do not correspond exactly (generally $\chi_\Sigma^2<chi_K$). However, they are approximately proportional when considering them as a property of each triangle. We expect further increasing the number of levels for the computation of the number variance, thereby allowing us to get closer to the asymptotic regime, would bring the two values closer together.  Let us further compare the analytical results for the Veech triangles with the numerics. In general, we see the values from the numerics are slightly smaller than the analytical results. This is not entirely unexpected, since the analytical periodic orbit calculations take into account only the diagonal approximation \cite{bogomolnyPeriodicOrbitsContribution2001} and higher orders in the asymptotic expansion and contributions from diffractive orbits are not analytically available. The triangles $V_5$ and $V_7$ consistently give similar values of the compressibility (and have similar spectral statistics in general), while  the numerics show quite distinct spectral statistics for $V_{10}$ and $V_{18}$.


\section{Conclusions and Discussion}\label{sec:Conclusion} 

The paper presents  the detailed analysis of the spectral statistics 
 of high excited energy levels of eight rational triangular quantum billiards belonging to the class of pseudo-integrable dynamical systems.  The principal  result is that the level spacing statistics of such billiards are of the intermediate type and  rather  accurately described by the model of gamma distributions. 
 
 The characteristic  features of observed statistics are: (i) the level spacing distributions exhibit level repulsion and have exponential tails. (ii) The shape parameter of the higher order level spacings $\gamma_n$ is linearly dependent on the order (iii) The level compressibility is non-trivial $0<\chi<1$, as seen in the linear regime of the number variance and the spectral form factor at the origin. (iv) The compressibility and proportionality coefficient of the level spacing shape parameter are  approximately  related by $\chi=1/p$. 
 
The normalized gamma distributions \eqref{eq:gammaP} provide a nearly perfect fit to the level spacing data. Thus, one can interpret them as Wigner-type surmise that is a quite close approximation to the underlying analytically unknown probability distribution. The linear relation of the shape parameters and order of the level spacings \eqref{eq:BrodyP} is consistent with previous results from structured random matrices and short-range plasma models \cite{bogomolnyStatisticalPropertiesStructured2021}, however the slope is not universal for all triangles. Since the level spacing distributions are easy to compute in contrast to the spectral form factors (or even the number variance) this provides an easily accessible way of determining the compressibility directly from the slope. The numerical results for the spectral form factor show a good agreement with the analytical expression \eqref{eq:LaplaceFun} derived from the gamma distributions. The results for the more general obtuse triangles show that the special lattice property of the Veech triangles seems to be  not essential for the applicability of the gamma model which may indicate that  it  is relevant for general  pseudo-integrable systems. 


\section{Acknowledgements}
We thank Tomaž Prosen and Barbara Dietz for inspiring discussions, as well as the organizers of the 8th Dynamics Days Central Asia and Caucasus conference in Bukhara for facilitating the opportunity for the authors of this paper to meet in person. Č. L. thanks the Max Planck Society for its hospitality. 

\appendix
\section{Toy model}\label{AppendixA}
In this paper it has been observed that the simple gamma fit \eqref{eq:gammaP}
describes the $n^{\mathrm{th}}$ nearest-neighbor distributions for all considered triangular billiards quite well.  The choice of such fitting function is  not based  on profound theoretical grounds. It is just  a simple function which  tends to zero at small  $x$ to describe the level repulsion and has an exponential tail at large argument, which is a characteristic feature of the intermediate statistics. In addition, there exists models where such type of functions is exact \cite{bogomolnySpectralStatisticsQuantum2004}. 
   
To understand  the accuracy of gamma fits, it is instructive to consider the following toy model. Let $M$ be a $2\times 2$ random symmetric  matrix 
\begin{equation}
M=\left (\begin{array}{cc} e& v\\v&-e  \end{array}\right )
\end{equation}
where $e>0$ and $v>0$  are real random variables with probability densities  $R(e)$ and $Q(v)$
\begin{equation}
\int_0^{\infty}R(e)de=1,\qquad \int_0^{\infty}Q(v)dv=1. 
\end{equation} 
Notice that the distributions are normalized over the interval $[0,\infty)$. When distributions are symmetric, this is a matter of convention.

The eigenvalues of this matrix are $\lambda_{1,2}=\pm \sqrt{e^2+v^2}$ and the  distribution of the spacing is 
\begin{multline}
P(s)=\int \delta(s-2\sqrt{e^2+v^2})R(e)Q(v)de dv=\\
=\frac{s}{4} \int_0^{\pi/2}  R(s\cos(\phi)/2) Q(s\sin(\phi)/2)d\phi\, . 
\label{general_formula}
\end{multline} 
When $R(e)$ and $Q(v)$ are  the Gaussians with zero mean and equal variance, one gets the usual GOE Wigner surmise. 

\begin{figure}
  \centering
  \includegraphics[]{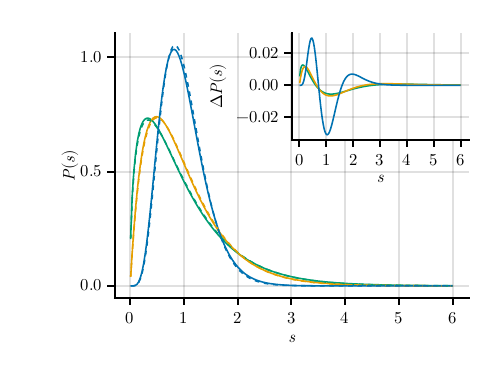}
  \caption{Toy model distribution \eqref{eq:toy_model} for different values of $\nu$. Green line: $\nu=-1/2$, yellow line: $\nu=0$, blue line: $\nu=4$.  Dashed lines of the same color indicate the gamma distribution \eqref{eq:gammaP} with $\gamma=1/2,\;1,\;5$, respectively. All distributions are normalized on the unit first momentum. Insert: the differences between the toy model distributions and the corresponding gamma distributions.} 
  \label{toy_model}
\end{figure}

Let us assume that $R(e)$ is the exponential function (to mimic the Poisson distribution) but off-diagonal variable $v$ is distributed according to  a gamma distribution $Q(v)$
\begin{equation}
P(e)=\exp(-e),\qquad Q(v)=\frac{v^{\nu} }{\Gamma(\nu+1)}\exp(-v)
\end{equation}
After a rescaling,  the spacing distribution in such a case  is 
\begin{widetext}

\begin{equation}
P^{(\mathrm{toy \;model})}(s)= \frac{\lambda^{\nu+2}  s^{\nu+1}}{2^{\nu+2} \Gamma(\nu+1)}\int_0^{\pi/2} \sin^{\nu}(\phi)\exp\left ( -\frac{\lambda s}{2}(\cos(\phi)+\sin(\phi)\right )d\phi
\label{eq:toy_model}
\end{equation}

where  $\lambda$ is fixed by the normalization of the first moment
\begin{eqnarray}
(n+1)\lambda&=&\int_0^{\infty} ds \frac{s^{\nu+2}}{2^{\nu+2} \Gamma(\nu+1)}\int_0^{\pi/2} \sin^{\nu}(\phi)\exp\left ( -\frac{s}{2}(\cos(\phi)+\sin(\phi)\right )d\phi\nonumber\\
&=&2(\nu+2)(\nu+1)\int_0^{\pi/2}\frac{\sin^{\nu}(\phi)d\phi}{(\cos(\phi)+\sin(\phi))^{\nu+3}}\, .
\end{eqnarray}

\end{widetext}
These integrals, in general, cannot be expressed in terms of known functions. In Fig.~\ref{toy_model} this distribution computed numerically  is plotted by solid color curves for three different values of $\nu=-1/2,\, 0,\, 4$ . The dashed lines of the corresponding color indicate the gamma distributions \eqref{eq:gammaP} with $\gamma=\nu+1$, i.e., which have the same power of $s$ at small argument as $P^{(\mathrm{toy \;model})}(s)$.  It is clearly seen that these simple formulas are in a good agreement with numerical calculated distributions  without any fits.  For $\nu=0$ this fact was mentioned in \cite{kongQuantumDynamicalTunneling2024}.   

The gamma distributions are not exact and  small deviations are, of course, present. In the Insert of  Fig.~\ref{toy_model} the differences between  $P^{(\mathrm{toy \;model})}(s)$ and the indicated gamma distributions are plotted. For small $\nu$ the difference is  of the order of $0.01$ but  for  $\nu=4$ it is around $0.03$. To get a better  approximation,  it is natural to fit  the parameter $\gamma$ in \eqref{eq:gammaP}  from the data and/or  to propose another fitting distributions. 

\begin{figure}
  \centering
  \includegraphics[]{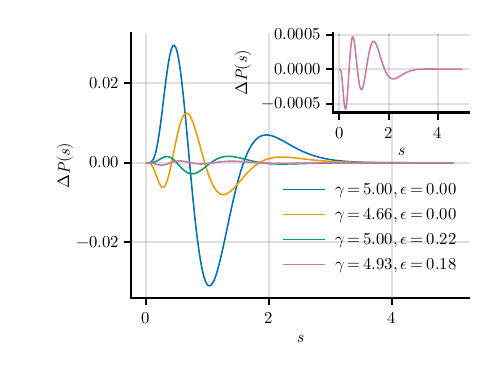}
  \caption{Differences between toy model distribution \eqref{eq:toy_model} with $\nu=4$ and different fits $\Delta P(s) = P^{\mathrm{(toy \;model)}}(s)-P^{\mathrm{(model)}}(s)$. Blue line: the gamma distribution with $\gamma=5$. Yellow line: the gamma distribution with fitted $\gamma=4.660$. Green line: the first correction to the gamma distribution \eqref{first_correction} with $\gamma=5$ and $\epsilon=0.218$. Magenta line: two parameter fit of \eqref{first_correction} with $\gamma=4.934$ and $\epsilon=0.181$.  Insert: the same magenta line as in the main figure but in a finer scale.} 
  \label{expanded_model}
\end{figure}

At Fig.~\ref{expanded_model} the differences between  $P^{(\mathrm{toy \;model})}(s)$  with $\nu=4$ and different fits are shown.  The blue line is the same as in the Insert of Fig.~\ref{toy_model} where the gamma distribution with $\gamma=5$ was used. The yellow line corresponds to the gamma distribution \eqref{eq:gammaP} but with the fitted value of $\gamma$.  It is the same procedure which used in the main text.  The fit gives  $\gamma\approx 4.66$.  The  use of a fitted $\gamma$ roughly speaking reduces the discrepancy twice with respect to $\gamma=5$. Though such simple approximation is enough for practical purposes, one may be interested in better estimations. The usual way of fitting unknown distributions consists in expanding them  in a series of suitable functions. As the gamma distribution by itself is a good first approximation one can, e.g., use the following series
\begin{equation}
P(s)=P^{(\mathrm{gamma})}(s)\left (1+\sum_{j=2} \epsilon_j p_j(s) \right )
\end{equation}
where $\epsilon_j$ are arbitrary constants and $p_j(s)$ are polynomials of order $j$ orthogonal with respect to $P^{(\mathrm{gamma})}(s)$
\begin{equation}
\int_0^{\infty} P^{(\mathrm{gamma})}(s)p_j(s)  p_k(s)ds=h_j \delta_{jk}. 
\end{equation}
For  $P^{(\mathrm{gamma})}(s)$ such polynomials are known as the generalized Laguerre polynomials (see e.g. \cite{batemanHigher1953}).  The sum starts with the polynomial of the second degree to get the imposed normalizations unchanged. The first correction term corresponds to the following expression,
\begin{widetext}

\begin{equation}
P(s)=\frac{\lambda^{\gamma+1}}{\Gamma(\gamma+1)} s^{\gamma} e^{-\lambda s}\left ( 1+ \epsilon \Big (1-\frac{2\lambda s}{\gamma+1}+\frac{\lambda ^2 s^2}{(\gamma+2)(\gamma+1)}  \Big ) \right ).
\label{first_correction}
\end{equation} 

\end{widetext}
This function depends on two independent parameters $\gamma$ and $\epsilon$. In Fig.~\ref{expanded_model} the yellow line is the difference between the numerically computed  toy model distribution with $\nu=4$  and function  \eqref{first_correction} with fixed $\gamma=5$ and fitted $\epsilon$. The one-parameter fit gives $\epsilon\approx 0.218$. The amplitude of remaining deviations is smaller than $0.003$.  One can also fit the data by  \eqref{first_correction} considered as  a function of two fitted parameters $\gamma$ and $\epsilon$. The fit gives $\gamma=4.934$ and $\epsilon=0.181$. The difference from the toy model distribution with these two parameters fitted is indicated in Fig.~\ref{expanded_model} by the magenta line. This difference is so small (less than $0.0004$) that it is  almost invisible in the scale of the figure. The same difference but at much finer scale is shown in the Insert of the same figure.

 Of course, one could invent many different formulas to fit an unknown distribution. Beside many of them, a simple  gamma fit gives a rather accurate approximation (in many cases less than  $0.01$) and has an additional advantage that all its moments can easily be calculated in a closed form. If necessary, higher order approximations can be used to find a more refined approximation.

\begin{figure*}
  \centering
  \includegraphics[]{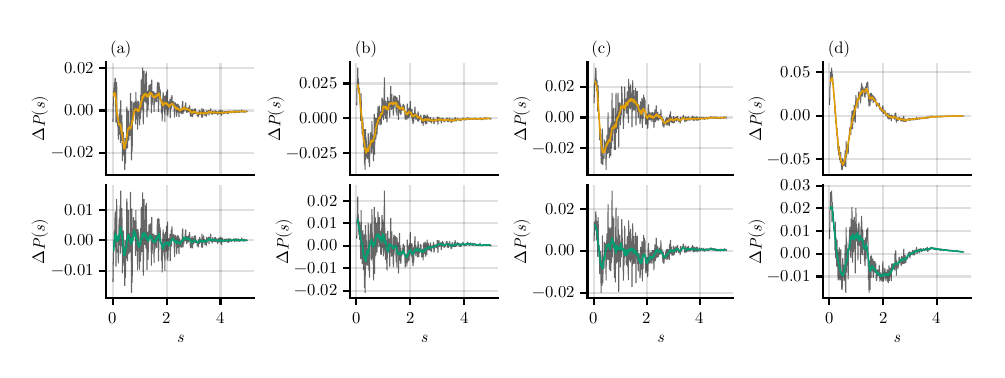}
  \caption{Comparison of the differences of the nearest neighbor level spacing distributions $\Delta P(s) = P^{\mathrm{(data)}}(s)-P^{\mathrm{(model)}}(s)$ for several triangular billiards and the refined two parameter gamma models using the correction given by Eq. \eqref{first_correction}. The tow row shows just the best fitted gamma model without correction terms, setting $\epsilon=0$ and the bottom row shows the correction with fitted $\epsilon$. The data is taken from the spectra of triangles (a) $V_{10}$, (b) $V_{18}$, (c) $T_1$, (d) $T_2$. The colored lines show the averages over 10 points to more easily identify the trends in the data.} 
  \label{expanded_model_data}
\end{figure*}

\begin{figure*}
  \centering
  \includegraphics[]{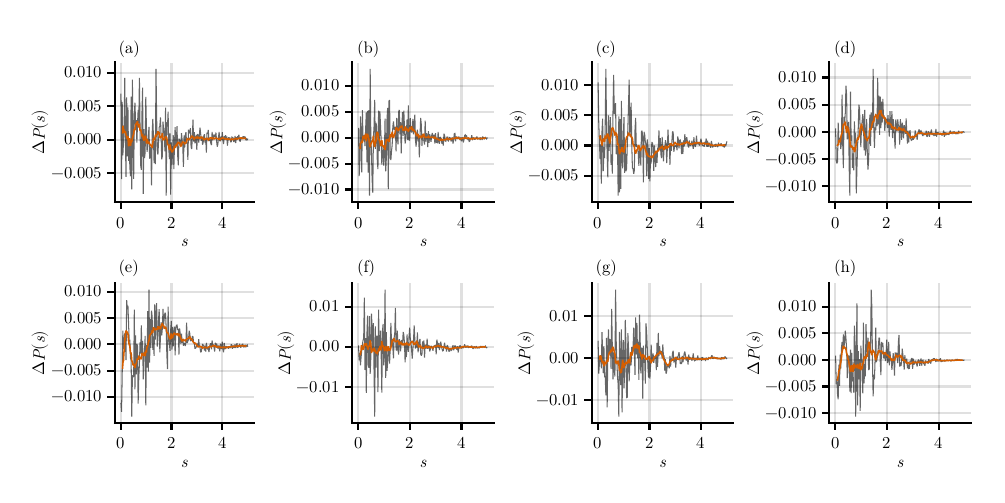}
  \caption{Comparison of the differences of the nearest neighbor level spacing distributions $\Delta P(s) = P^{\mathrm{(data)}}(s)-P^{\mathrm{(Bessel fit)}}(s)$ for all triangular billiards and the two parameter Bessel models using the correction given by Eq. \eqref{Bessel}.  The panels show the triangles (a) $V_5$, (b) $V_7$, (c) $V_8$, (d) $V_9$, (e) $V_{10}$, (f) $V_{18}$, (g) $T_1$, (h) $T_2$. The colored lines show the averages over 10 points to more easily identify the trends in the data.} 
  \label{bessel_model_data}
\end{figure*}

\begin{table}[]

\centering\caption{Best fitting parameters for the expanded gamma model \eqref{first_correction} and Bessel fit \eqref{Bessel}, fitted to the nearest neighbor level spacing distributions of the triangular billiards.}{\label{tab:fitting}}
\begin{tabular}{>{\centering}p{1.5cm}>{\centering}p{1.5cm}>{\centering}p{1.5cm}>{\centering}p{1.5cm}>{\centering}p{1.5cm}}
\multicolumn{5}{p{8.6cm}}{}\tabularnewline
\hline 
\hline 
label & $\gamma$ & $\epsilon$ & $\alpha$ & $\beta$  \tabularnewline
\hline 
\hline 
$V_5$ & 0.854 &  0.012 &  1.309 &  0.42 \tabularnewline
$V_7$ & 1.064 &  -0.057 &  1.825 &  0.892\tabularnewline
$V_8$ & 0.848 &  0.029 &  1.243 &  0.289  \tabularnewline
$V_9$ & 1.147 &  -0.132 &  2.366 &  1.447  \tabularnewline
$V_{10}$ & 0.848 &  -0.069 &  1.64 &  0.903 \tabularnewline
$V_{18}$ & 1.131 &  -0.123 &  2.314 &  1.409 \tabularnewline
$T_1$ & 1.225 &  -0.121 &  2.4 &  1.416 \tabularnewline
$T_2$ & 1.300 &  -0.315 &  4.64 &  3.688 \tabularnewline

\hline 
\end{tabular}
\label{tab:two_parameter}
\end{table}

\section{Two parameters fits}\label{AppendixB}

It is plain that better approximations can also be developed for  level distributions of triangular billiards. Consider, for example, the nearest-neighbor distributions for the right triangles with angles $\pi/10$ and $\pi/18$. In Fig.~\ref{expanded_model_data} the differences between numerical data and the gamma fits are presented in the top row (panels (a) and (b)). Though the gamma fits  give quite good results, small regular deviations  (similar to the toy model) are clearly visible at these figures. The same may be seen for the two obtuse triangles (panels (c) and (d)).  The bottom row shows the two parameter fit \eqref{first_correction}  with fitted values of $\gamma$ and $\epsilon$. This helps to reduce the difference between the data. The same is true for the obtuse triangles presented in panels (c) and (d). The fit is especially good for the triangle $V_{10}$ (see panel (a)) as it seems to be structureless and random.

Let us consider, for diversity,  a different kind of two parameter fit
\begin{equation}
P_n^{(\mathrm{Bessel\;fit})}(s)=c\, q^{\alpha+1}  \,s^{\alpha} \,K_{\beta}\big (q s \big )
\label{Bessel}
\end{equation} 
where $K_{\beta}(x)$ is the modified Bessel function of the third kind. Constants  $c\equiv c(\alpha,\beta)$,  and $q\equiv q(\alpha, \beta,n)$  are  determined from the normalization conditions
\begin{widetext}
\begin{eqnarray}
c(\alpha, \beta)&=&\left (  2^{\alpha-1} \Gamma((\alpha-\beta+1)/2) \Gamma((\alpha+\beta+1)/2) \right )^{-1},\\ 
q(\alpha, \beta,n)&=&\dfrac{2 \Gamma((\alpha-\beta)/2+1) \Gamma((\alpha+\beta)/2+1)}{(n+1) \Gamma((\alpha-\beta+1)/2) \Gamma((\alpha+\beta+1)/2)}. 
\end{eqnarray}
\end{widetext}

This function has two independent parameters: $\alpha$ and $\beta$. Such a function but with $\beta=\alpha-1$ has been proposed in \cite{gonzalezUnifiedStochasticFormaism2024} to approximate the nearest-neighbor distribution for a quantum lima\c{c}on billiard with mixed phase space. 
 
Following the known asymptotic behaviors of $K_{\beta}(x)$, the Bessel fit \eqref{Bessel} has power asymptotics at small $s$ and exponential at large $s$. But contrary to the gamma fit, the powers at small and large $s$ may be different. For  half-integer $\beta$ $K_{\beta}(x)$ is expressed through elementary functions. In particular, for $\beta=1/2$ the Bessel fit coincides with the gamma fit. In Fig.~\ref{bessel_model_data} the difference between the nearest-neighbor level spacing distributions for all triangles and the Bessel fits are shown. The resulting oscillations are small and in some cases appear purely random. As expected, two parameter fits give better approximations but are more complicated and less transparent. The best fitting parameters are presented in Table \ref{tab:two_parameter}.

\FloatBarrier
\bibliography{references} 

\end{document}